
\vsize=22.5cm
\hsize=14.5cm
\def\sk { \vskip .6cm }
\def\capit#1 {\vfill \eject \sk \centerline {\nbc #1}  \vskip 1cm }
\def\newpar#1 { \vskip 1.5cm \line {\nbb #1 \hfill}  \vskip .4cm}
\tolerance=1600

\overfullrule=0pt
$  $
\vskip1.5truein
\centerline {\bf Stabilizing the gravitational action}
\centerline {\bf and Coleman's solution}
\centerline {\bf to the cosmological constant problem}
\bigskip
\centerline{Alberto Carlini$^{(1)}$ and Maurizio Martellini$^{(2)}$}
\smallskip
\centerline{\it (1) International School for Advanced Studies, SISSA,}
\centerline{\it Strada Costiera 11, I-34014 Trieste, Italy}
\centerline{\it (2) Department of Physics, University of Milan,}
\centerline{\it Via Celoria 16, I-20133, Milan, Italy}
\centerline{\it and}
\centerline{\it I.N.F.N., University of Pavia, Pavia, Italy}
\vskip1truein
\centerline {\bf Abstract}
We use the 5-th time action formalism introduced by Halpern and Greensite
to stabilize the unbounded Euclidean 4-D gravity in two simple minisuperspace
models.
In particular, we show that, at the semiclassical level ($\hbar \rightarrow
0$),
we still have as a leading saddle point the $S^4$ solution and the Coleman peak
at zero cosmological constant, for a fixed De Witt supermetric.
At the quantum (one-loop) level the scalar gravitational
modes give a positive semi-definite Hessian contribution to the 5-D
partition function, thus removing the Polchinski phase ambiguity.
\vskip2truein
\vfil
\eject
{\bf 1. Introduction.}
\medskip
As it is well known, the Euclidean theories with classical action unbounded
from below have a Boltzmann factor $\exp (-S)$ which is not normalizable in
the partition functional integral $Z_E$.
This integral is not well defined, and the theory is unstable against large
field fluctuations.

To avoid these difficulties, some authors have introduced a scheme borrowed
from stochastic quantization [1].
Their main idea was to construct a stabilized theory where the
Boltzmann factor $\exp (-S)$ is substituted by a normalizable factor $\exp
(-S_{eff})$ which has the same classical limit as that of the bottomless
theory.
In particular, for the wrong-sign $-\lambda \phi^4$ theory and for the large
$N$ field expansion, the two actions $S$ and $S_{eff}$ have the same
perturbative expansion, but this is not the case for gravity [6].

These results can be achieved because the instabilities of the bottomless
theory appear to be nonperturbative effects.
The basic idea is to imagine that the starting D-dimensional theory is, in
fact, due to an underlying D+1-dimensional quantum theory.
Recently, this ``D+1-th time'' formalism has been used in order to provide a
truly non-perturbative definition of 2-D quantum gravity [2].

In this letter we shall show that this formalism also allows to solve the
Polchinski's phase ambiguity [3] found in Coleman's solution [4] to the
cosmological constant problem.
Namely, we shall show how to stabilize the Euclidean 4-D quantum gravity
action,
which is unbounded from below, so that to have again Coleman's mechanism for
the vanishing of the cosmological constant without any phase ambiguity in the
``stabilized'' Euclidean path integral.
This is possible since, at the semiclassical level (i.e. $\hbar \rightarrow
0$),
the effective action of the stabilized theory is the same as that one coming
from the standard bottomless 4-D gravity action.
Thus we still have as leading saddle point a $S^4$ ansatz solution and the
Coleman peak at the vanishing cosmological constant, for a given choice of the
De Witt supermetric (see section 4).
Moreover, already at the one-loop approximation, the stabilizing
term starts to contribute so that the Weyl (scalar) modes of the gravitational
(metric) field give now a positive semi-definite Hessian contribution to the
partition function of the stabilized theory.
Remember that the phase ambiguity in the path integral of the bottomless
Einstein action is just due to these scalar modes [3].
\bigskip
{\bf 2. Stabilizing bottomless action theories: a review.}
\medskip
The main idea of ref. [1] is to consider the vacuum expectation value $<Q>$ of
an operator $Q$ (depending
on a set of fields $\phi$) as the expectation value in the ground state (GS)
$\Psi_0$ of the D+1-dimensional theory :
$$
<Q>={1\over Z_E}\int ~d\phi~e^{-S/\hbar}~Q[\phi]=<\Psi_0\vert Q\vert
\Psi_0>~~~,
{}~~~~~\Psi_0[\phi]={1\over \sqrt {Z_E}}e^{-S/2\hbar}\eqno(2.1)
$$
The D+1-dimensional theory is defined by the Hamiltonian $H$ for which $\Psi_0$
is the GS.
The Hamiltonian is :
$$
H_{D+1}=\int ~d^Dx~~\left [-{1\over 2}{\delta^2\over \delta \phi^2}+{1\over 8
\hbar^2}\left ({\delta S\over \delta \phi}\right )^2-{1\over 4\hbar}{\delta^2
S\over \delta \phi^2}\right ]\geq 0\eqno(2.2)
$$
Now, since $\Psi_0$ given by eq. (2.1) is well defined (normalizable) only for
bounded theories, while $H_{D+1}$ is positive semi-definite (and then it has a
well
defined GS for any $S$), the strategy is to assume the D+1-dimensional theory
with $H_{D+1}$ as the more fundamental theory.

The general method is to look for the GS of the Fokker-Planck equation :
$$
H_{D+1}\Psi_0=E_0\Psi_0~~~,~~~~~\Psi_0={1\over \sqrt {Z}}e^{-S_{eff}/2\hbar}
\eqno(2.3)
$$
and to redefine eq. (2.1) with this new $\Psi_0$ ($S_{eff}$ is the effective
D+1
action defined through $H_{D+1}$).
It is obvious that, if the starting action is bounded, $S_{eff}=S$ and $E_0=0$.
But for a bottomless action one will have, in general, $S_{eff}\not =S$ and
$E_0>0$, and the new theory is normalizable and stable.
Since, in most cases, it will be extremely difficult to analytically solve
eq. (2.3), one can define the theory in terms of a path integral formulation,
where one has (in D=4) :
$$
<Q>=~{1\over Z_5}\int ~d\phi (x, x_5)~Q[\phi (x, x_5=\bar x_5)]\cdot \exp
[-S_5]
\eqno(2.4a)$$
with
$$
S_5=~\int ~d^4xdx_5~\left [{1\over 2}(\partial_5\phi)^2+{1\over 8
\hbar^2}\left ({\delta S\over \delta \phi}\right )^2-{1\over 4\hbar}{\delta^2
S\over \delta \phi^2}\right ]\eqno(2.4b)
$$
and one computes expectation values on $x_5=const$ slices of the extra
dimension variable $x_5$.
Eq. (2.4) immediately says that, in the $\hbar \rightarrow 0$ limit,
the classical equation of motion for the bottomless action is recovered
(enforced), as it is required.
Moreover, the $O(\hbar )^{-1}$ term helps stabilizing the theory.
To see how this works, ref. [1] makes the examples of the wrong sign $-\lambda
\phi^4$ theory and the linearized 4-D quantum gravity.
\bigskip
{\bf 3. Stabilizing 4-D gravity.}
\medskip
Let us now try to specialize the ``machinery'' described in the previous
section to the case of bottomless gravity in 4 dimensions.
We will first consider the more general case of an expansion around a
minisuperspace ansatz for the original 4-D theory, and then we will study
the case of the de Sitter $S^4$ ansatz used by Coleman (C), discussing
the consequences on the objections raised by Polchinski [3].

The first step to generalize the previous results to the case of the
gravitational theory is to introduce the 5-D metric :
$$
ds_5^2=dt_5^2+ds_4^2\eqno(3.1a)
$$
$$
ds_4^2=N^2d\tau^2+a^2d\Omega^2_3\eqno(3.1b)
$$
where $t_5$ is the `fictitious' extra dimensional variable.
For simplicity we assume to work in the gauge $N=1$ (the ``proper time''
gauge for a closed synchronous minisuperspace ansatz), since the gauge-fixing
is actually unnecessary for perturbative calculations in the $S_5$ approach
[1]-
which is a sort of stochastic quantization [5].
In other words, one is left with the problem of fixing, at most, the four
dimensional diffeomorphisms.

The starting action for gravity will be taken as :
$$
S_G =  \int d^4x ~ \sqrt {g_4}\left (-{R\over k^2}+\lambda \right )
\eqno(3.2)
$$
where $k^2=16\pi G$.
For simplicity we consider the case of manifolds with compact topology, but
the results can be easily extended to the non-compact cases.

The natural extension of eq. (2.4b) to the Euclidean gravity, following
Greensite [6], leads to the action $S_5$, invariant under ($t_5$-independent)
4-D diffeomorphisms :
$$
\eqalignno {
S_5=& \int ~dt_5d^4x ~\biggl [{1\over 2k^4} G^{\mu \nu \alpha \beta} \partial_5
g_{\mu \nu} \partial_5 g_{\alpha \beta} +{k^4\over 8\hbar^2}G^{-1}_{\mu \nu
\alpha \beta} {\delta S_G \over \delta g_{\mu \nu}}\cdot {\delta S_G \over
\delta g_{\alpha \beta}}\cr
& -{k^4\over 4\hbar}G^{-1}_{\mu \nu \alpha \beta}
{\delta^2 \over \delta g_{\mu \nu} \delta g_{\alpha \beta}}\biggr \vert_{order}
S_G\biggr ]&(3.3)\cr}
$$
where $g_{\mu \nu}$ is the 4-D metric given by eq. (3.1b) (which then
substitutes $\phi$ as the functional integration field in eq. (2.4a)) and the
only change from [6] accounts for a different choice of the dimensional factors
of $k^2$.
$[...]\vert_{order}$ reflects the arbitrariness in the possible choice
of the operator ordering of the supermetric and functional derivatives, and
$G^{\mu \nu \alpha \beta}$ is the De Witt supermetric :
$$
G^{\mu \nu \alpha \beta}={1\over 2} \sqrt {g}\left (g^{\mu \alpha}g^{\nu
\beta}+
g^{\mu \beta}g^{\nu \alpha}+c g^{\mu \nu}g^{\alpha \beta}\right )
$$
$$
G_{\mu \nu \alpha \beta}={1\over 2 \sqrt {g}}\left (g_{\mu \alpha}g_{\nu
\beta}+
g_{\mu \beta}g_{\nu \alpha}-{c\over 1+2c} g_{\mu \nu}g_{\alpha \beta}\right)
$$
where $c$ is an arbitrary parameter ($c> -{1\over 2}$ for the positivity of
$G^{\mu \nu \alpha \beta}$ and for the stability of $S_5$).

Performing the functional derivatives in eq. (3.3) for the simple case $c=0$
(which does not restrict the generality of the results), and assuming
the order written, we end with the following expression for the 5-D action:
$$
\eqalignno {
S_{5}=&2\pi^2\int ~d\tau dt_5~\biggl [{6a\over k^4}\left ({\partial a\over
\partial t_5}\right )^2+{1\over 8a\hbar^2} [12(\ddot a^2 a^2+\dot a^4
+1-2\dot a^2 +\ddot a\dot a^2 a-a\ddot a)\cr
& +6\lambda k^2(\ddot a a^3+\dot a^2 a^2-a^2)+\lambda^2 k^4 a^4 ]
+{1\over \hbar}a^3\lambda k^4\delta^4_{inv}(0)\biggr ]&(3.4)\cr}
$$
where $\delta^4_{inv}$ is the 4-D diffeomorphism invariant delta.

Now we will consider the effect, on such an action, of the quantum fluctuation
$\phi$
of the field $a$ around a background classical configuration $\bar a$ :
$$
a(t_5,\tau)=\bar a(t_5,\tau)+k^2\hbar \phi (\tau)~~~, ~~~~~\bar a=const
\eqno(3.5)
$$
Here we take the quantum fluctuations $\phi (\tau)$ depending on $\tau$ only,
which is a sort of ``crude'' notation for the idea that the gauge group in
the stabilized gravity theory is formed only by $t_5$-independent
diffeomorphisms.
For simplicity (but this assumption does not restrict the generality of our
one-loop results ), we consider the case where $\bar a=const $ and therefore
we expand in $k^2$ the action (3.4) around $\bar a$ as :
$$
S_{5}=(2\pi^2)\int ~dt_5d\tau ~[A(\hbar)+k^2\hbar B(\hbar)\phi +k^4\hbar^2
\phi M(\hbar)\phi +O(k^6)]\eqno(3.6)
$$
The first term of the right hand side of this equation is just the classical
contribution to $S_{5}$, and the second term can be put equal to zero by
means of a suitable redefinition of $\phi$.
Then, at the one-loop approximation, the only important quantum contribution
to $S_{5}$ (as it is well known,
see, e.g., [7]) is the third, Gaussian term of eq. (3.6).
Its explicit expression is :
$$
\hbar^2M={3k^4\bar a\over 8}\left [4\left ({d^2\over
d\tau^2}+{1\over \bar a^2}\right )^2+2\lambda k^2 {d^2\over d\tau^2}+
\lambda k^4 (\lambda +8\hbar \delta_{inv}^4(0))\right ]\eqno(3.7)
$$

As it is well known, the Gaussian fluctuation in the quantum field $\phi$
gives rise to a one-loop determinant of the effective theory :
$$
\int ~d\phi ~e^{-k^4\hbar^2\phi M\phi}\simeq [Det ~(k^4\hbar^2M)]^{-1/2}
\eqno(3.8)
$$
The theory is well behaved if the $M$ operator has no negative or zero
eigenvalues [8].
To the lowest order in $\hbar$, the spectrum of $\hbar^2M$ is given by :
$$
\hbar^2M\rightarrow {3k^4\bar a\over 2}\left ({d^2\over d\tau^2} +{1\over \bar
a^2}\right )^2~>0,~~~~if~~~~\lambda k^2\rightarrow 0\eqno(3.9)
$$
So, in this simple case we have shown that the prescription of
considering the 5-D effective action is really a working `trick' which allows
to
stabilize the theory against the quantum field fluctuations.

The next step would be then to consider the more general case of a 4 metric
with an arbitrary lapse function $N$ and to tackle the discussion of the
Fokker-Planck equation (2.3), in particular to find the true ground state.
{}From a technical point of view, this should amount to solve the Wheeler-de
Witt
functional equation associated to the Fokker-Planck Hamiltonian (2.2), for
instance using the minisuperspace approximation [9] and the operator ordering
corresponding to the proper time gauge choice [10], which is the simplest one.
This is not a simple task and it will be the matter for another future paper
[11].
\bigskip
{\bf 4. The C-mechanism revisited. }
\medskip
We are now in the position to analyze the behaviour of the C-theory for the
cosmological constant $\lambda$.
The fundamental interest in this discussion is motivated by the observations
made by Polchinski [3], Hawking [12], Unruh [13] and Veneziano [14].
Polchinski claimed that the double exponential in the C-partition function
($P\sim \exp \left (\exp \left ({1\over \lambda }\right )\right )$), describing
an infinite set of wormhole connected universes, is {\it not} peaked at
$\lambda
=0$.
In fact, the determinant of the quantum fluctuations of the conformal part
of the metric field produces, at one-loop, an overall phase $(i)^{D+2}$, which
in D=4 dimensions turns the original distribution into a disappointing $P\sim
\exp \left (-\exp \left ({1\over \lambda}\right )\right )$, destroying
the C-model.
Actually, this observation is tightly correlated to the more general
interpretation of [12, 13], which consider the C-divergence at $\lambda =0$
as the indirect reflection (a `reminiscence') of the {\it unboundedness} of
the Euclidean action for gravity, and not necessarily due to topological
effects.

Using the method of the 5-th time action, we will show below the interesting
fact that {\it the claimed peak at $\lambda =0$
is still present in the stabilized effective
theory, in the classical limit, with no `disrupting' effects due to the
one-loop
quantum fluctuations}.
The conclusion is that one should revalue Coleman results as something more
intrinsic and fundamental in their own.

To analize in details the C-mechanism, we will consider a metric of the form :
$$
ds^2_5=dt_5^2+ds^2_4\eqno(4.1a)
$$
$$
ds^2_4=r^2d\Omega_4^2\eqno(4.1b)
$$
where $r$ is the radius of the 4 sphere $S^4$.
The action for gravity is :
$$
S_G(r)=-\int ~d^4x~\sqrt {g_4}\left ({R\over k^2}-\lambda\right )={8\pi^2
\over 3}\left (\lambda r^4-6{r^2\over k^2}\right )\eqno(4.2)
$$
and the Ricci scalar curvature is $R={12\over r^2}$.
Thus, using the previous formulas, we get that
the 5-D effective action is given by :
$$
\eqalignno {
S_{5}=&{8\pi^2\over 3}\int ~dt_5~\biggl [{8(1+2c)\over k^4}r^2\left ({\partial
r\over
\partial t_5}\right )^2+{k^4\over 8\hbar^2(1+2c)}\left ({6\over r^2k^2}-\lambda
\right )^2r^4\cr
& +{\lambda k^4\over 4\hbar}{(4+9c)\over (1+2c)}r^4\delta^4_{inv}(0)
\biggr ]&(4.3)\cr}
$$
Already at a first naive inspection, this formula suggests that the classical
C-solution $r^2={6\over k^2\lambda}$, which is enforced in the $\hbar
\rightarrow 0$ limit by the effective action (4.3), and for which the starting
4-D action is unbounded from below ($S_G\simeq -{3\over \lambda k^4}\rightarrow
_{\lambda
\rightarrow 0}~-\infty $), is actually stabilized by the term $-{\delta^2S_G
\over \delta r^2}\rightarrow {1\over \lambda}{(4+9c)\over (1+2c)}
\rightarrow_{\lambda\rightarrow 0}+\infty$ (for $c>-{4\over 9}$).

To see how this mechanism works in a more precise way, we first need knowing
which are the classical solutions ($\bar r$) for the effective action $S_{5}$.
To study these solutions, we once again assume to work in the ansatz $\bar r=
const$, and then we easily find the following equation of motion :
$$
{\delta S_{5}\over \delta r}={8\pi^2\over 3\hbar^2}{\lambda k^4\over (1+2c)}
\int~dt_5~r^3\left [{1\over 2}\left (\lambda -{6\over k^2r^2}\right )
+\hbar(4+9c)\delta_{inv}^4(0)\right ]\biggr\vert_{\bar r}=0\eqno(4.4)
$$
which admits the classical ($\hbar \rightarrow 0$) solutions :
$$
\lambda \bar r^2={6\over k^2},~~~S_{5}={24\pi^2\over \hbar}{(4+9c)\over (1+2c)}
\int dt_5{\delta_{inv}^4(0)\over \lambda}\eqno(4.5a)
$$
$$
\bar r=0,~~~S_{5}=0~~(trivial)\eqno(4.5b)
$$
It is then evident that the 4-D C-solution survives (eq. (4.5a) also in this
``enlarged'' theory, and it mainteins its singular character ($S_{5}
\rightarrow -\infty$), provided $\lambda \rightarrow 0^+$ and we choose $c\in
(-{1\over 2}, -{4\over 9})$, or $\lambda \rightarrow 0^-$ and $c>-{4\over 9}$.

We now study the behaviour of the quantum fluctuations around these
classical solutions.
If we put :
$$
r(t_5, \tau)=\bar r+k^2\hbar\theta(\tau)\eqno(4.6)
$$
we find that $S_{5}$ can be expanded (to the second order in $\theta$) as :
$$
S_{5}={8\pi^2\over 3}\int ~dt_5~[A^{\prime}+k^2\hbar B^{\prime}\theta +
k^4\hbar^2\theta M^{\prime}\theta +O(k^6)]\eqno(4.7)
$$
where $A^{\prime}=A^{\prime}(\hbar , \bar r)$, $B^{\prime}=B^{\prime}(\hbar ,
\bar r)$ and $M^{\prime}=M^{\prime}(\hbar , \bar r)$.
As before, the interesting contribution at the one-loop approximation comes
from the Gaussian term :
$$
\hbar^2M^{\prime}={3\over 2}{\lambda k^8\over (1+2c)}\left [{1\over 2}\left
(\lambda \bar r^2-{2\over k^2}\right )+\hbar \bar r^2(4+9c)\delta_{inv}^4(0)
\right ]\eqno(4.8)
$$
In particular, we find that, for the classical solution corresponding to the
C-model ($\bar r^2={6\over \lambda k^2}$) :
$$
\hbar^2M^{\prime}\simeq {3\lambda k^6\over (1+2c)}>0~~~,~~for~~\lambda \bar
r^2={6\over k^2}\eqno(4.9)
$$
(to the lowest order in $\hbar$) which gives a well defined stable partition
function, with {\it no phases at all}.
This result suggests that the Coleman solution to
the cosmological constant problem should survive in the context of the
``fifth-time'' Einstein action
and that it should not be the mere consequence of the fact that the 4-D
gravitational action for the $S^4$ sphere is negative definite and hence
unbounded from below [12, 13].
\bigskip
{\bf Acknowledgements.}
\medskip
We would like to thank Prof. G. Veneziano for useful discussions, and the
Referee of Phys. Lett. B for stimulating us to better clarify some points.
This work has been supported by the Italian
Ministero per l'Universita' e la Ricerca Scientifica
e Tecnologica.
\vfil
\eject

\def\tindent#1{\indent\llap{#1}\ignorespaces}
\def\ref{\par\hang\tindent}


{\bf References:}
\medskip
\ref{$1.$ }J. Greensite and M.B. Halpern, {\it Nucl. Phys. } {\bf B242}, 167,
(1984).
\ref{$2.$ }J. Ambjorn, J. Greensite and S. Varsted, {\it NBI} preprint
{\it HE-90-39}, (1990).
\ref{$3.$ }J. Polchinski, {\it Phys. Lett. } {\bf B219}, 251, (1989).
\ref{$4.$ }S. Coleman, {\it Nucl. Phys.} {\bf B310}, 643, (1988).
\ref{$5.$ }G. Parisi and Wu Yong-Shi, {\it Sci. Sin.} , {\bf 24}, 483, (1981).
\ref{$6.$ }J. Greensite, {\it Nucl. Phys.} {\bf B361}, 729, (1991).
\ref{$7.$ }P. Ramond, {\it ``Field Theory, a modern primer''}, (ed. The
Benjamin/Cummings Publ. Comp., 1981).
\ref{$8.$ }G.W. Gibbons, S.W. Hawking and M.J. Perry, {\it Nucl. Phys.}
{\bf B138}, 141, (1978).
\ref{$9.$ }J. Hartle and S.W. Hawking, {\it Phys. Rev.} {\bf D28}, 2960,
(1983).
\ref{$10.$ }C. Teitelboim, {\it Phys. Rev. } {\bf D28}, 297, (1983).
\ref{$11.$ }A. Carlini and M. Martellini, {\it SISSA} preprint, in preparation.
\ref{$12.$ }S.W. Hawking, {\it Mod. Phys. Lett.} {\bf A5}, 453, (1990).
\ref{$13.$ }W.G. Unruh, {\it Phys. Rev.} {\bf D40}, 1053, (1989).
\ref{$14.$ }G. Veneziano, {\it Mod. Phys. Lett.} {\bf A4}, 695, (1989).
\vskip1.5truein
\vfill
\eject
\bye